\documentclass[dvips,a4paper,10pt]{article}
\usepackage{dcolumn}% Align table columns on decimal point
\usepackage[pdftex]{graphicx}
\usepackage{float,rotating,subfigure}
     \usepackage[small,sc]{caption}
\usepackage{float}
\usepackage{physics}
\usepackage{geometry} 
\graphicspath{{./Pictures/}} % Specifies the directory where pictures are stored
\usepackage{amsmath}
\usepackage{mathtools}
\usepackage{amssymb}
\usepackage{amscd}
\usepackage{listings}
\usepackage{amssymb}
\usepackage{amsthm}
\usepackage{amsfonts}
\title{An experimental Scheme for Gravitational Scattering in Microscale:\\
\large{The effect of spatial superposition of mass on the microstructure of space-time}}
\date{\today}
\author
{Sahar Sahebdivan,$^{1\ast}$\\\\
\normalsize{$^{1}$University of Vienna, Boltzmanngasse 5, 1090 Vienna, 
Austria}\\
\\
\normalsize{$^\ast$ E-mail:  saharsahebdivan@univie.ac.at \& saharsahebdivan@gmail.com}
}

\begin{document}

\maketitle

\begin{abstract}

In this paper, we are exploring the feasibility of observing non-classical features of gravity in a low-energy regime in a quantum optomechanical experiment. 
If gravity is to have an underlying quantum nature, it should hold the most fundamental quantum characteristics such as the superposition principle and entanglement. Despite the weakness of gravity, in principle there is a chance, to observe such a quantum signature of the gravity by exploiting the quantum optomechanical techniques, without direct observation of graviton. 
We are investigating a new dynamical scheme called, gravitational quantum regime, in which the source of gravity is a quantum particle, and its centre of mass is subject to the spatial superposition. 
In a Gedankenexperiment, a test particle is gravitationally interacting with a quantum nanoparticle in a double-slit setup.  Possible entanglement or superposition of the fields is investigated. 
We are looking for the corresponding deviation of the classical description of gravity despite being far from the Planck scale. Any experimental interrogation which reveals that gravitational field obeys the quantum superposition principle would be the first recognition of quantumness of gravity.
This study will show how feasible it is to search for a non-classical feature of gravity in such a regime of motion.
Moreover, this proposal would be an attempt to test the objectivity of the quantum superposition principle and its contribution to the microstructure of space-time. 

\end{abstract}
\section{Motivation}
\subsection{Introduction}
Spatial quantum superposition of a massive object is one of the most intriguing ideas of quantum mechanics. The notion conflicts the common sense of causality, and as we will see, if it comes to the question on how such an object might gravitate, it contradicts the classical concept of space-time. 
In particular, knowing the exact value and behaviour of the gravitational field of a particle in spatial superposition, known as a Spatial Cat State (SCS), is a subtle task - even theoretically.
This theoretical challenge -which had been propounded first by Feynman- has influenced the pioneering attempts to unify quantum theory with gravity and was taken as the basis for whether or not gravity should be quantised like the other fundamental force fields \cite{feynman2003, eppley1977necessity}. 

Ventures surrounding the experimental realisation of the Spatial Cat State for massive objects has received significant interest due to the recent developments in experimental physics and has been the subject of many current investments
\cite{pino2018chip,palomaki2013entangling,ulbricht2018testing,carlesso2017cavendish}.  Due to the sensitivity of the experiment to the environmental parameters, like noises, space-based operations are planned to be launched via satellites to test it in the microgravity condition \cite{kaltenbaek2016macroscopic}.

The paradigm of these recent investigations on the SCS is to study physical systems in the interface between gravity and quantum mechanics where both theories have a non-negligible contribution to the dynamics of the system. The goal is to, first, test the extent and limitation of the quantum superposition principle, and second, examine different arguments and theories both in favour of and against the quantisation of gravity.
Following a scientific debate on the necessity of quantising gravity \cite{albers2008measurement, carlip2008quantum, mattingly2005quantum}, few proposals have been initiated to certify or falsify the quantisability of the gravitational field experimentally \cite{bose2017spin, marletto2017gravitationally, derakhshani2016probing, pino2018chip}. However, most of these proposals are highly model-dependent and have been built upon specific assumptions or interpretations. For example, two very similar proposals suggested by Bose. et.al \cite{bose2017spin} and Marletto. et al. \cite{marletto2017gravitationally, marletto2018can} on ``Gravity induced entanglement'', assume a specific form of quantum-classical interaction to certify the quantised nature of the gravitational field as a quantum intermediary for entanglement. 
As argued in \cite{hall2018two, Diosi2015spontaneous}, presuming a particular model of gravitational interaction \cite{kafri2015bounds, kafri2014classical} introduces a loophole in a final analysis on the nature of gravity. 

Another example, is the gravitational collapse model introduced by Penrose and Di{\'o}si \cite{penrose1996, Diosi2007notes}, which dispute the quantisability of gravity by predicting a ``gravity-related collapse ''  \cite{tilloy2016sourcing, derakhshani2016probing, bahrami2014schrodinger} -are based on hypothetical theories of gravity. Moreover, any gravity-related collapse model assents an objective interpretation of quantum mechanics. 

To circumvent problems associated with this model-dependency \cite{hall2018two, Diosi2015spontaneous}, it is essential to propose a phenomenological approach with minimal presumptions regarding the mechanism of gravitational interaction. 
In this paper, we try to fulfil this need by introducing a schematic experiment to measure gravity 
in microscale and when the source mass is in the quantum regime of motion. 
We begin with a very general proposition that states if gravity has an underlying quantum mechanism, there should be an associated observable which, directly or indirectly, adheres to the principles of quantum theory. We also assume that the Newtonian limit of linearised gravity is valid for low ranges of energy. As there are strong evidences to show the Spatial Cat State is realisable in practice for matter-waves \cite{arndt2014testing}, it is reasonable to ask whether the spatial coordinate of the centre of mass can be subject to the quantum superposition.
Hence, we search for a probable effect of such a spatial superposition on the gravitational field of the object, by looking for a deviation from the inverse square law. 
The scale of such effect (if it exists) is expected to be larger than any other quantum gravity effects. Such a deviation would be different and independent from those predictions of string theory or any other quantum gravity model, as the underlying cause would be different in the order of magnitude and nature. \cite{bjerrum2003quantum, bertolami2003ultracold,ford1993semiclassical}. 

The well-recognised complexity in unifying quantum theory and gravity comes from the fact that gravitational fields do not behave as smoothly as other fundamental forces under the process of quantisation. However, this is not the only area of the incompatibility between the two theories \cite{sabin2010, sabinart, Amelino2013}.
When external degrees of freedom, coordinate position, is subject to the superposition principle,
as, in case of in SCS, there is a conflict with the classical concept of gravitational -or electrostatic- potential \cite{ford1982gravitational, ford1993semiclassical}. 
In other words, to calculate the gravitational field of a single particle in an SCS, we need to associate a quantum superposition of two solutions of a unique Schr\"odinger equation, with
a classical superposition of two solutions of two different Poisson equations. The only way around this difficulty is to interpret the system statistically. Nevertheless, in section (\ref{sec:2}), we will show that the statistical interpretation of the formalism fails to describe a particular example of a double slit experiment with single particles. However, before that, we need to review some crucial aspects of the double slit experiment and probabilistic expression of repeated and individual events in quantum mechanics. 

\subsection{Double slits experiment: Is there still an unanswered question?}

The most convincing evidence of the experimental realisability of a Spatial Cat State (SCS), is a double-slit experiment with matter-waves: an archetype of quantum mechanics. 
According to Feynman, the behaviour of particles in a double slit is the basic or even the only, mystery of quantum mechanics \cite{feynman1965feynman}. In fact, the history of modern physics is full of various Gedanken-and real- experiments with a double slit setup, each revealing a specific perspective or shedding light on less explored angles of the quantum reality. 
Among many examples we may recall the Heisenberg microscope \cite{schrodinger1930heisenberg}, Einstein-Bohr recoiling double slit \cite{bohr1928quantum}, WheelerÕs delayed-choice \cite{ma2013quantum}, Scully {\it{et.al.}} complementarity measurement \cite{scully1991quantum}, Eppley-Hannah's experiments \cite{eppley1977necessity} and many more.
Despite that, after many decades of progress in understanding quantum physics, we still have enough reason to believe that the mystery remains. 

The formation of the interference pattern of particles has been examined and observed repeatedly over the last half-century. The experiment has been performed with fundamental particles as well as large molecules, and continues to be reproduced with more massive objects \cite{arndt2014testing, pino2018chip, schmole2016micromechanical} and larger separation distance.
Whether or not there is a limit on how large an object can be to get into a spatial superposition is one of the main questions in GQR studies.  
 
An SCS is a spatial superposition of the centre-of-mass of a nano- or micro-object:
\begin{equation}
\label{eq:cat_state}
    \ket{\psi}=
    \frac{\ket{x_{1}}+\ket{x_{2}}} {\sqrt{2}}.
\end{equation} 
The quantum state, $\ket{\psi}$, is represented in terms of position eigenstates and is a prototype of a pure quantum state with maximal entanglement. 
Although a pure Cat State, such as a two level-system, is one of the foremost commonly used expressions of quantum mechanics, it becomes intricate to describe an individual quantum object in an SCS -which is essential in testing gravity.  
 
\begin{figure}
\center
\hspace*{0 cm}{\includegraphics[width=0.5\textwidth]{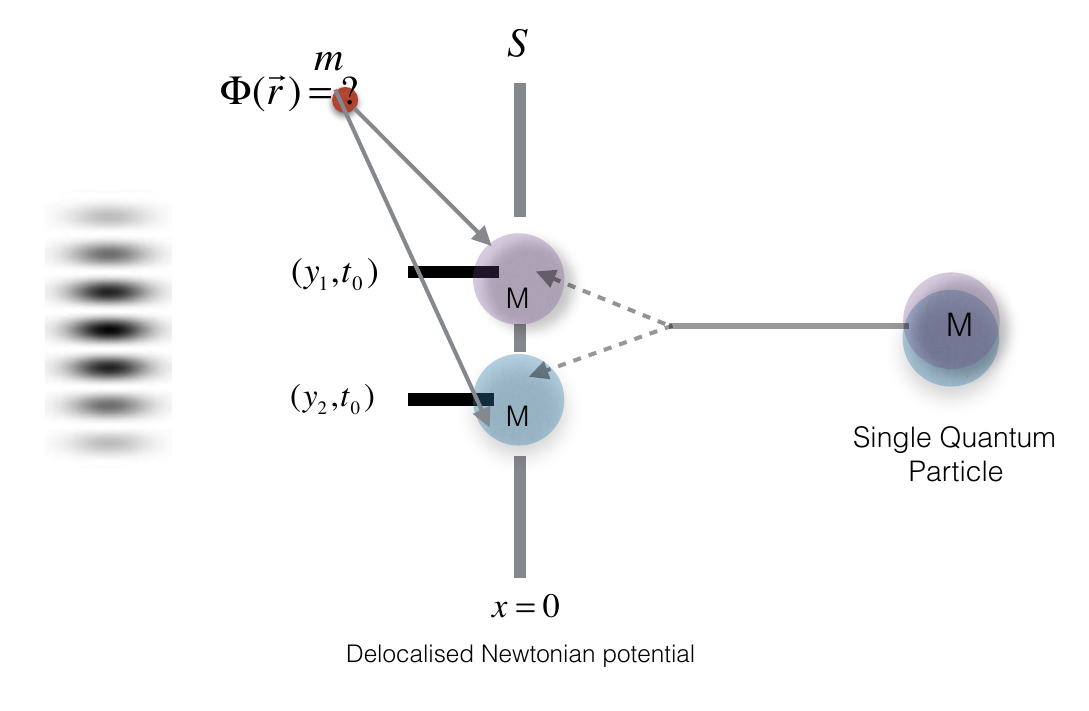}}
\caption{}
\label{fig:double-slits}
\end{figure}

Schematically, a double slit experiment is constructed from a stream of particles emitted toward a barrier with two or more narrow windows {\textbf{{\textbf{fig.}}}}(\ref{fig:double-slits}). The windows constrain the position of incoming elements and results in the diffraction of particles on the other side of the barrier. The stream of the particle is prepared coherently and sent from a source or oven toward the barrier. If the coherency conditions are fulfilled, the famous interference fringes are formed- even for massive particles- at the screen. The experiment is considered for {\it{single particle}} if each element arrives at the barrier one at a time with sufficiently large spatial and temporal intervals \cite{juffmann2012real}.
The practical fact that we are neither able to predict theoretically nor observe experimentally the path that each individual particle takes still is at a great oddity. 
Any measurement that tries to determine the trajectory of each particle would result in losing the fringes of the interference pattern. It is believed that the exchange of momentum due to measurement leads to the destruction of quantum correlations and therefore, the loss of interference \cite{scully1991quantum} {\footnote{
Even in the case of weak measurement, still, there is a trade off efficiency for accuracy, which results in a change in the interference pattern. This outcome evokes the crucial role of measurement in quantum mechanics: any observational outcomes semantically and quantitatively has a strong dependency on the process of measurement \cite{mir2007double}.}}.
This behaviour is interpreted as the complementarity between particle and wave nature of each quantum entity, and as Bohr manifested, is the direct consequence of the uncertainty principle.
Nevertheless, despite such a fundamental inability, by reading the repeated outcomes of detectors, we can construct a wavefunction and determine the quantum state, SCS, at the location of slits barrier, which describes the process of interference.
The flair that brings us around this problem is hidden in the intrinsic statistical nature of quantum mechanics. If we consider the path, from the source to the screen, taken by each particle as a single instance, the formation of interference is the result of numerous instances.  
In this manner, we are able to explain the system of many single particles statistically; hence the wavefunction has an intrinsic statistical nature.

Still, there is a problem: what is the wavefunction for each individual particle? Isn't quantum mechanics capable of saying anything about the state of each particle at the slits? 
The conventional interpretation of quantum mechanics instructs us that going from repeated (or collective) behaviour of a system to the single event is equal to going from statistics to probability. 
Therefore, from the perspective of the standard interpretation of quantum mechanics for particular observable, we can assign a probability to each individual particle.
Accepting the statistical/probabilistic nature of the formalism solves all the mystery of the double slit experiment. 
Extrapolating the probability of a single event from the statistical behaviour of the collective (or repeated) instances is not limited to quantum mechanics, however, expressing an intrinsic probabilistic nature to an observable {\it{is}} a peculiarity of quantum mechanics. 
Following this picture, one can conclude that the {\textbf{eq.}}(\ref{eq:cat_state}) should also represent an individual particle as many authors widely accept it. 
This notion means an individual massive quantum particle can occupy more than a unique position in the time, which is in contrast with classical physics.

However, we claim that none of the conventional double slit interferometry experiments, in reality, have examined whether an SCS, {\textbf{eq.}}(\ref{eq:cat_state}), does, in fact, represent a single individual particle, or if it is only a statistical extrapolation of repeated numerous instances. One reason for such a vacuum could be the extreme technical difficulty of isolating and controlling an individual quantum object. 

While the matter of individuality is redundant for interferometry or the {\it{welcher weg}} type experiments \cite{durr1998origin}, it becomes significant if we want to measure the dynamical properties such as gravity. 
If the wavefunction is a statistical representation of many repeated events, other observables of the system, like position or energy, would be assigned probabilistically. 
However, it is not well-defined to associate a dynamical quantity like a gravitational field to such a state. 

On the other hand, if the SCS is valid for an individual quantum object, we should be able to assign the gravitational property to an SCS. The validity of SCS for an individual object is interpreted in the literature as self-interference, delocalisation or ``happening to occupy'' two spatial coordinates simultaneously, which is in contrast with classical physics. 
There is no exclusive answer to this question beyond different opinions. Nevertheless, we believe that answering this question would have a significant consequence in quantum physics.
In the appendix A., we summarise the two pictures of the wavefunction, while both can be categorised as standard Copenhagen interpretation, each might lead to different experimental results. 

If the statistical picture is correct, the Spatial Superposition State would also be a statistical semantic or a representation, probably, without any causal consequence. Therefore, any SCS experiment would not reveal any extra information on the foundation of physics, particularly intrinsic collapse models. Accordingly, there would be no obvious implication on gravity in this type of experiments. 

Contrarily, if the SCS is a valid quantum state for a single particle, there would be a breakdown in the causal structure of classical physics and particularly space-time. Such a breakdown, in principle, would be in the table-top range of energies and if detected experimentally, can fundamentally change our understanding of space-time on the micro-level. Therefore we see an indispensable need to have a new paradigm in wave-matter interferometry which asks the proper question. 

Our proposal is going to examine such a hypothesis, by assuming the second case: A single quantum particle can be in a state of Spatial Superposition and therefore be delocalised over a spatial interval due to its inherent uncertainty in dynamical variables. If so, there would be a correction to the inverse square law of gravity. Such a correction can be read from the interference pattern of the interferometry experiment. 
For that, we are going to see how we can describe a single quantum particle through a double slit setup. 

\section{Self-interference of a single quantum event} \label{sec:2}

In the previous section, we have said that if the SCS, {\textbf{eq.}}(\ref{eq:cat_state}), which represents an SCS, would be a valid state of a single, individual particle, there would be consequential conflicts with the classical concept of a force field. The conflict is more crucial when it comes to gravity- compared to electromagnetism for example; One reason is that in the case of electromagnetism the field is quantised and that, to some degree, eases the conflict. And more importantly, gravity, in the classical picture, is the cornerstone of physical causality and space-time, and anything that conflicts that picture would be of a greater issue. In this section, we will see the difficulty in analysing the gravitational field of the particle right at the moment that it goes through either or both slits {\textbf{{\textbf{fig.}}}}(\ref{fig:double-slits}).

In a double slit experiment, we are working with quantum particles such as atoms, large molecules, and even nano-objects, which are often highly localised and non-spreading. The mathematical expression of such components is a non-trivial task in quantum mechanics. While they are in the quantum regime of motion, neither the Schr\"odinger equation nor Dirac fields can properly express the dynamics of such entities{\footnote{ Solutions of the Schr\"odinger equation are spreading by nature, while particles in double slit interferometry experiments are fairly localised.}}. 
To the best knowledge of this author, the only theoretical framework that allows describing an individual, localised, rigid (non-spreading) quantum object is the de Broglie wavelet model introduced by O. A. Barut, in 1993, \cite{barut1993diffraction}. However, in this stage we focus on the SCS
which are solutions of Schr\"odinger equation, we show by applying the standard, non-relativistic quantum mechanics it fails to describe the gravitational potential of an a spatialy superposed object.

\subsection{Gravitational potential of an object in a double slit}

In this part, we are going to calculate the gravitational field of particle $M$ in a double slit experiment. For now, we assume that the gravitational field of a quantum particle follows the Newtonian law of gravity. The setup is confined to the $x-y$ plane as shown in the {\textbf{fig.}}(\ref{fig:double-slits}). The solid barrier $S$ is located at $x=0$, and contain two narrow slits that restrict the position of the particle to ${y_{1}}$ and ${y_{2}}$. Therefore we can consider the barrier $S$ as a measuring position apparatus which can be represented as a quantum operator $\hat{S}$ with two Eigenstates $\ket{y_{1}}$ and $\ket{y_{2}}$. Particle $M$ comes from a source at $x \rightarrow -\infty$ and passes through the slits. As we discussed before an SCS, describe such a particle 
$ \ket{\psi}= \frac{1}{\sqrt{2}}({\ket{y_{1}}+\ket{y_{2}}})$.
To know the value of an observable $\hat{A}$ of the particle $M$ in the state $\ket{\psi}$, we need to calculate the expectation value of the operator.
First, let us recall the way that we observe the position of the particle at either of slits in a {\it{welcher weg}} experiment. 
 In a {\it{welcher weg}} experiment, we detect the passage through each slits. Operator $\hat{A}$, measure the location of passing particle at $x=0$, the expectation value of such a measurement would be:

\begin{equation}
\label{eq:operator_A}
\bra{\psi}\hat{A}\ket{\psi} = 
    \frac{1}{2}\left(\bra{y_{2}}+\bra{y_{1}}\hat{A}\ket{y_{1}}+\ket{y_{2}}\right)=
    \frac{1}{2} \left(\bra{y_{1}}\hat{A}\ket{y_{1}}+\bra{y_{2}}\hat{A}\ket{y_{2}}\right)
\end{equation}

Apparatus $\hat{A}$, which can be a detector, is a different position measuring operator than barrier $\hat{S}$, however, it measures the same observable, and hence, it commutes with $\hat{S}$. In this particular case, both operators have two mutual Eigenvalues {\footnote{The Hilbert space spanned by Eigenstates of operator $\hat{S}$ is the subspace of the Hilbert space of operator $\hat{A}$}}:

\begin{equation}
\hat{A} \ket{y_{i}} = y_{1}\ket{a_{i}}
\end{equation}

Therefore, {\textbf{eq.}}(\ref{eq:operator_A}) would be reduced to: 

\begin{equation}
\bra{\psi}\hat{A}\ket{\psi} = 
    \frac{1}{2} \left({y_{1}^{2}}+{y_{2}^{2}}\right)
\end{equation}

As we can see the interference terms disappears. After achieving the ``welcher weg'' information, only one path would be taken, and there are no quantum superposition \cite{storey1994path}.  
Now, let us change the setup. We want to know the gravitational field at of $M$ the point $\vec{r}$ in the region $x > 0$ at the moment that particle passes through the barrier without knowing the path that it takes.

The probability distribution of the location of source mass $M$ at $x=0$ is:
\begin{equation}
\label{eq:prob}
    {\mathcal{P}}(y)_{|x=0} = 
        |\psi(x,y)_{|x=0}|^{2} = 
       |\bra{y}\ket{\psi}|^{2} =
            \frac{ |\bra{y}\ket{y_{1}}|^{2} + |\bra{y}\ket{y_{2}}|^{2} }  {2} 
         + \bra{y}\ket{y_{1}}\bra{y}\ket{y_{2}}
\end{equation}

In our assumption, the gravitational field still is not a quantum observable, and for the moment, we are not going to define the quantum operator for the gravitational field observable systematically. But still we can consider the following realistic toy model based on standard quantum mechanics: to measure the gravitational field of the particle $M$, we place another test particle $m << M$ at point $\vec{r}$. Particle $m$ is just located at its position to ``feel'' the gravitational effects of particle $M$. It is safe to consider the gravitational recoil on $M$ is negligible. We describe the ``feeling '' of the particle $m$ at point $\vec{r}$ received from $M$ is an SCS, by the following model:

\begin{equation}
\label{eq:field}
    {g(\vec{r})}=\frac{{\mathcal{G}} M} {|\hat{r} |^{2}},
\end{equation}
where ${\mathcal{G}}$ is a gravitational constant.
We can assume that the operator $\hat{r}$ in {\textbf{eq.}}(\ref{eq:field}) is also a sort of position measuring apparatus, which performs a weak measurement on the position of $M$.
If we repeat the same process of finding expected value that we've had for detector $\hat{A}$, we will see:

\begin{equation}
\label{eq:operator_m}
\bra{\psi}\frac{1}{\hat{r}^{2}}\ket{\psi} = 
    \frac{1}{2}\left(\bra{y_{1}}+\bra{y_{2}}\frac{1}{\hat{r}^{2}}\ket{y_{1}}+\ket{y_{2}}\right)
    \end{equation}

There is a fundamental differentiation  between eq.(\ref{eq:operator_m}) and eq.(\ref{eq:operator_A}) that would create a complexity: 
Operator $\frac{1}{\hat{r}^{2}}$ does not have naturally, a set of discrete Eigenstates, while $\hat{S}$ has. As both are still measuring position, both operators should commute, but they don't share the same Hilbert space of Eigenstates.
Eigenfunctions of the operator $\hat{r}$ are continuous and not quadratically integrable.
But considering the physical functionality of $\hat{r}$, and we can define the following relations:

\begin{align}
\label{eq:r}
     \hat{r} \ket{r} &= r \ket{r}  \\
     \hat{S} \ket{y_{i}}_{s} &= y_{i}\ket{y_{i}}_{s} \\
     \hat{r} \ket{y_{i}}_{s} &= (r -y_{i})\ket{y_{i}} = (r -y_{i}) \delta ({y_{i}})
\end{align}

and consequently: 

\begin{equation}
\label{eq:inverse}
\frac{1}{\hat{r}^{2}} \ket{y_{i}}_{s} = \frac{1}{|r^{2}-y_{i}^{2}|}\ket{y_{i}}
\end{equation}
For operators with continuous Eigenstates, the Born rule would be written in the integral form of:

\begin{equation}
\label{eq:continious}
    \langle \hat{r} \rangle = \int_{\mathbb{R}} r \: |\psi(r)|^{2} \: \mathrm{d}r
\end{equation}

If the particle $M$ goes with probability ${\mathcal{P}}(y_{i})$ through the slit at $y_{i}$, then, from eq.(\ref{eq:continious}) and eq.(\ref{eq:prob}), the probability of finding the value of the field at point $\vec{r}$ would be:

\begin{equation}
\label{eq:gravity1}
     \langle{g(\vec{r_{0}})} \rangle=
     {\mathcal{G}} M \int_{\mathbb{R}^{2}}
        \left(
     \frac{1}  {\hat{r}^{2}} {\mathcal{P}}(y_{1}) +
     \frac{1}  {\hat{r}^{2}} {\mathcal{P}}(y_{2}) +
     \frac{1}  {\hat{r}^{2}}  \bra{y}\ket{y_{1}}\bra{y}\ket{y_{2}} \right) {\mathrm{d}}r^{2}
\end{equation}
By applying eq.(\ref{eq:inverse}) on eq.(\ref{eq:gravity1}), we will have:

\begin{eqnarray}
 \langle{g(\vec{r_{0}})} \rangle = 
    & {\mathcal{G}} M {\mathcal{A}} \int_{\mathbb{R}^{2}} 
    \left( 
         \frac{\delta(r-{y_{1}}) \delta(r-{y_{1}})}  {|{r_{0}} - (r - {y_{1}})|^{2}} + 
         \frac{\delta(r-{y_{1}}) \delta(r-{y_{1}})}  {|{r_{0}} - (r - {y_{2}})|^{2}} +
         \frac{\delta(r-{y_{1}}) \delta(r-{y_{2}})}  {|{r_{0}} - (r - {y_{1}})| 
                                          |{r_{0}} - (r - {y_{2}})|}
    \right) 
    {\mathrm{d}}^{2}r ,
\label{eq:gravity2}
\end{eqnarray}

Where ${\mathcal{A}}$ is the normalisation factor. As we can see, there are few problems with eq.(\ref{eq:gravity2}), particularly what is important to us, is the last term in the equation. The term is mathematically ill-defined and conceptually in an obvious contradiction with inverse square law. The root of this contradiction is that the inverse square law is highly local law of physics, the value of the field in one point of space only depends on the value of the charge in the other point position. Hence, there is no way to associate a gravitational field with a term like:

\begin{equation}
\label{eq:interface}
         \bra{y}\ket{y_{1}}\bra{y}\ket{y_{2}} = \psi(y_{1})\psi(y_{2})
\end{equation}

Such term implies a quantum superposition on the gravitational field, and that's only possible if the gravitational field would be a quantum field. This term, in the language of Feynman, is the most or even the only odd of quantum mechanics and the reason for his argument that gravity should be a quantised field.
The purpose of this model was only to show that calculating the gravitational field of a particle in an SCS is not a trivial task in standard non-relativistic quantum mechanics. 

The last term in {\textbf{eq.}}(\ref{eq:gravity1}), implies that the value of the field in position $\vec{r}$ does not just depend on the value of a field at position $y_{1}$ but also on the value at point $y_{2}$. This dependency is not an unfamiliar notion, but rather is a core concept in Kirchhoff diffraction theory: the value of the field at the screen does not depend only on the value of filed at slit $y_{i}$ but also depend on the relative phase (the optical path difference) with the other slit. This dependency is not that we expect for the gravitational (or even electrostatics) potential in classical physics. While the fields, either photon field, matter field, or any other fundamental fields which -consist of an enormous amount of quanta- can perform interference, it is odd when a single source which assumes to follow the inverse square law is the subject of the same interference mechanism. In classical physics, we do not face such a problem because
a source (gravitational or electrostatics) occupies a unique position in the space-time, and the field only depends on this unique coordinate. 

To go around this complicated situation, we either need to accept Feynman's argument that believes by considering the gravitational field as a quantum mediator this conflict would be solved.
Alternatively one might consider the Penrose- Di{\'o}si argument that an individual object cannot be in a Spatial Superposition State to save the classical structure of gravity. 

Inevitably there would be a third option that the inverse square law in microscale needs to be modified. As we will see in the next part, in the lack of rigorous theoretical framework, the final answer only will be achieved phenomenologically and experimentally \cite{Amelino2013,hooft2016good}. However, for that purpose, it is essential to have a deep understanding of the problem and asking the right key question. 

\subsection{Feynman and Penrose: two converse arguments}\label{sec:feynman_penrose}

We have seen that the gravitational field of a quantum particle in spatial superposition is in conflict with the classical concept of force fields.
Feynman addressed this concern many decades ago in his lecture notes on gravity \cite{feynman2003}. He argues that we are unable to answer the question regarding the potential associated with the term in eq.(\ref{eq:interface}) unless gravity is quantised as electromagnetism and other force fields.
This conclusion was the origin of the consistency argument for quantising the gravitational field. 
A decade later, in 1977, Eppley and Hannah improved the consistency argument is in favour of quantising of gravity, in the form of a thought experiment asserting to demonstrate that the ``gravitational field must be quantisedÓ \cite{eppley1977necessity}. Though, their argument endured several critiques on the validity of their deduction \cite{mattingly2006eppley,carlip2008quantum,mattingly2005quantum}. 
Inspired by Eppley-HannahÕs paper, in 2008, Albers, Kiefer and Reginatto, mathematically prove that no consistency argument can designate in favour or against the quantised nature of the gravitational field, but only an experimental verification can bring insights into this problem \cite{albers2008measurement}. In addition to that, Feynman failed to provide a description of why spatial superposition does not occur for macroscopic systems. His argument also fails to explain how a delocalised source of gravity would integrate into a relativistic model of space-time. Consequently, quantum gravity which is descendant of that approach is silent to the matter. In fact, even building a quantum theory of gravity with standard approaches, does not give means to address such problems. 
In the early nineties, Penrose and Di{\'o}si took the stand on the SCS, to opposing reason against the quantisability of the gravity. Such arguments lead to initiating a family of models known as gravitational collapse models. 
Penrose, in his 1996 paper \cite{penrose1996}, argues that the concept of spatial superposition is not a well-defined concept for a single quantum particle in the framework of non-relativistic quantum mechanics. Consequently, as we cannot have a spatial superposition for macro objects, we need a reason similar to super-selection rules and that might be gravity itself.
Penrose and Di{\'o}si have tried to prove that because of this question, gravity cannot be fundamentally quantised or if it is as soon as mass becomes large it decoherence (or collapse) into a classical regime \cite{tilloy2016sourcing}. Therefore, there should be a limit that beyond that the interference pattern would fade out. 
Begin with the semi-classical picture of gravity as the fundamental nature of space-time, conclude that there is a mass limit which beyond that any spatial superposition will collapse. 
This picture does not provide any argument about how would be the structure space-time below that limit. How a massive particle in spatial superposition would gravitate and how it affects the microscopic structure of space-time.

To summarise, we have seen that there is a legitimate question that has a fundamental role in our understanding of gravity in the quantum scale: whether an SCS is realisable for an individual particle, and if so what would be its gravitational field. We conclude that neither of the existing arguments or experiments could wholly and consistently answer this question without a thorough revise in current approaches. 
In the next section, we introduce an experimental scheme that can provide our purposes. Our proposal offers a means to analyse a different aspect of quantum coherence in gravity while avoiding pre-assumption about the nature or form of gravitational interaction.

\section{Experimental Scheme}

Based on arguments in this paper, we propose an experimental scheme to fulfil the promises, which are essential to solving one of the most long-lasting problems in quantum mechanics. The purpose is first to examine the objectivity of the superposition principle: whether an SCS is a valid state for an individual particle described by a de Broglie wavelet, and if so, determine and distinguish the gravitational field of an individual quantum particle in a Superposition Cat State.

\begin{figure}[h]
\begin{center}
\begin{subfigure}
%\resizebox{6in}{!}
%\hspace*{-2cm}
{\includegraphics[width=0.6\textwidth]{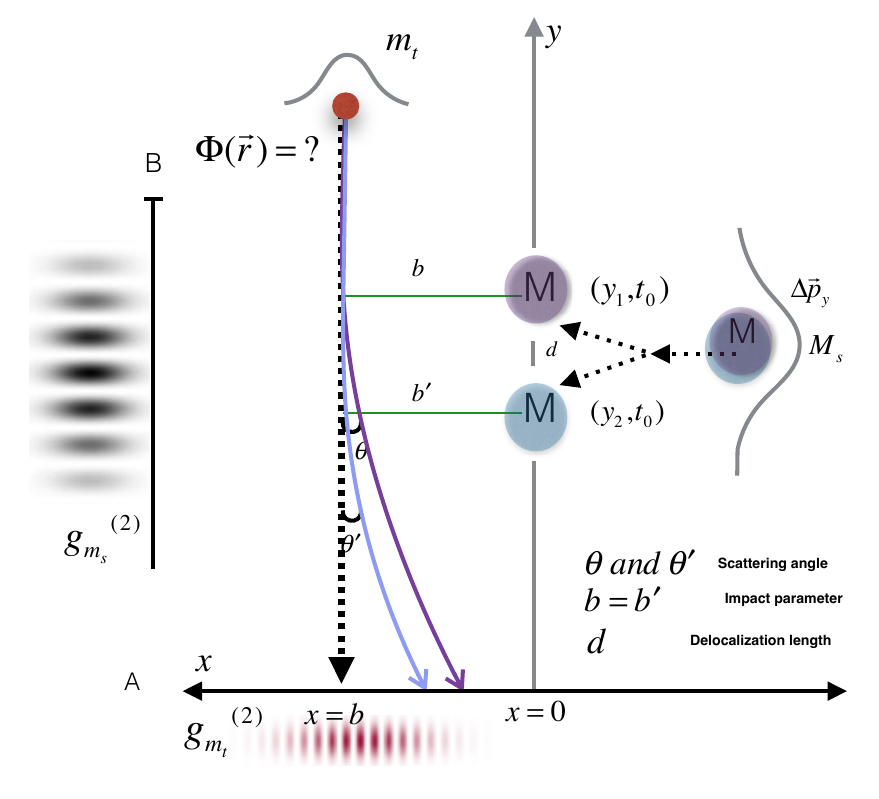}}
\caption{SCHEME OF THE MECHANISM: A quantum rigid (non-spreading) object $M$ is prepared in a Spatial Cat State (SCS). When $M$ passes through the slits, according to QM, its centre of mass delocalises of an order of $d$ to form the interferences. To examine the nature of the gravitational potential in such a state, a probe test mass $m_{t}$, is placed to scattered off the `` delocalised'' potential. Expected trajectories are showing (classical paths) if the mass $M$ were to go through either of the slits. $b$ is the impact parameter and $\theta/\theta'$ are deflecting angles. By reading out the interference patterns on screen $A$, associated with particle $m_{t}$, we can learn about  different regimes/conditions in which either entanglement or interference could form. This study gives an insight into the nature of the gravitational potential of particle $M$, and objectivity of spatial superposition of the centre of the mass.}
\label{fig:m42}
\end{subfigure}
\end{center}
\end{figure}

Our suggested scheme examines the gravitational scattering off a source mass in a spatial superposition state {\textbf{fig.}}(\ref{fig:m42}), and then provide the means to study the dynamical response of a probe mass $m$ to such a gravitational potential. The proposal consists of two components:\\

\begin{figure}[h]
\begin{center}
%\resizebox{6in}{!}
%\hspace*{-1cm}
{\includegraphics[width=0.7\textwidth]{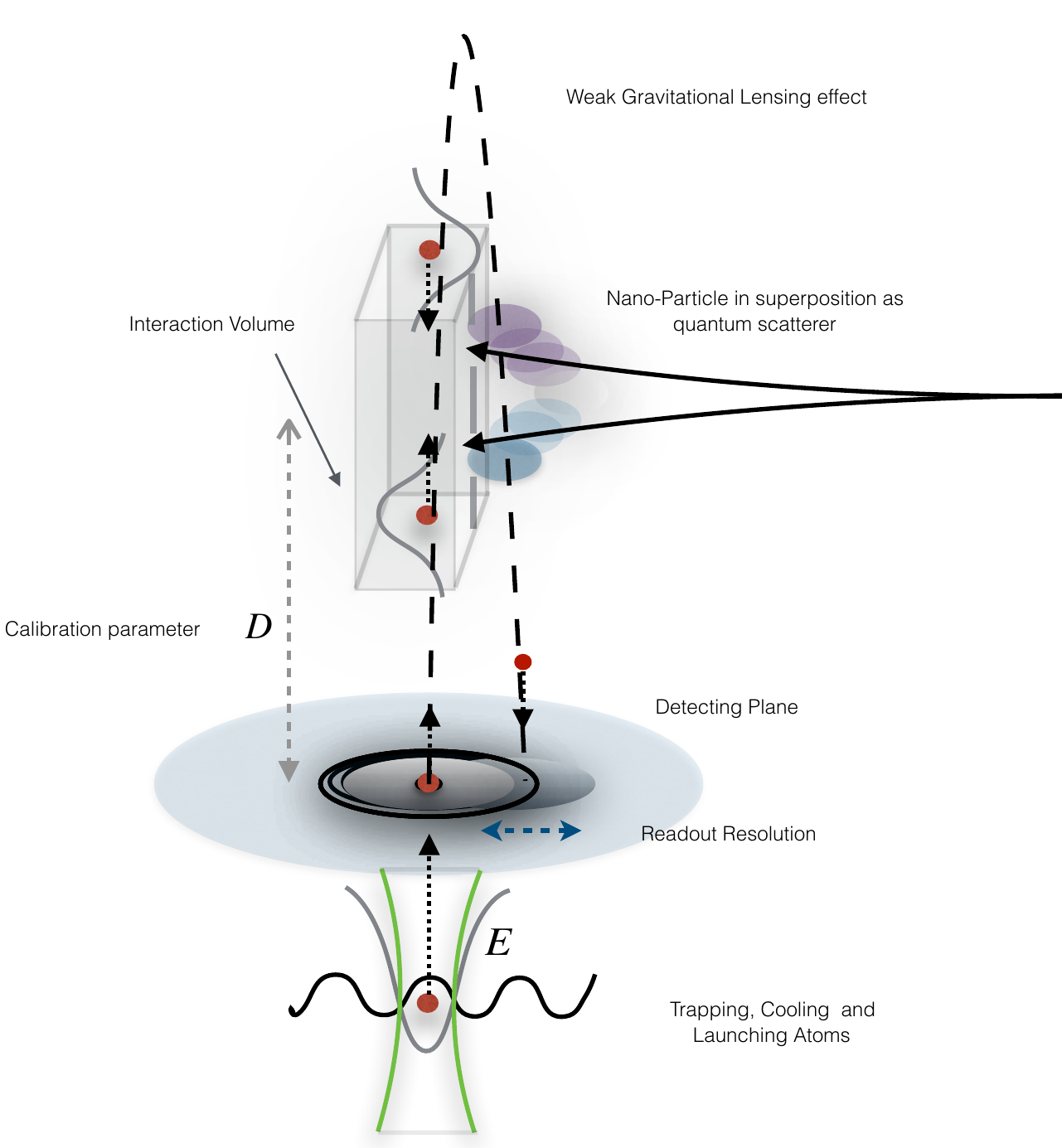}}
\end{center}
\caption{OPTIMISED MODEL OF EXPERIMENT: A quantum rigid mass $M$ is prepared in a Spatial Cat State (SCS) for a fraction of a second. While passing through the slits, its centre of mass delocalised as the order of $d$. The probe test mass $m_{t}$ is a fountain of cold atoms, cooled in an MOT system, launched upward to enter interaction volume. The missile shape of the setup is designed to function similar to gravitational lenses for matter wave, enhance the deflect angle and therefore the readout resolution. By reading out the displacement on the readout screen, we can learn about the nature of the gravitational potential of the source particle in an inverse scattering manner.}
\label{m44}
\end{figure}

\textbf{Scatterer}: a quantum object ($M$) is considered as the source of the gravitational field, and it is prepared in a Spatial Cat State via double-slit interferometry and injected into the cavity {\textbf{fig.}}(\ref{m44}) While passing through the slits, its centre of mass delocalised as the order of $d$ for a fraction of a second.\\

\textbf{Scattered Test Particle}: Test mass $m_{t}$ is used as a dynamical probe to study the response to the gravitational field of $M$, which is in an SCS. As it shows in the sketch of the experiment, {\textbf{fig.}}(\ref{fig:m42}), expected trajectories of the test mass $m_{t}$ (classical paths) are the indicator of, $M$ were to go through either of the slits. $b$ is the impact parameter and $\theta/\theta'$ are deflecting angles. By reading out the interference patterns on screen $A$, associated with particle $m_{t}$, we can learn about different regimes/conditions in which either entanglement or interference could form.

In a modified design {\textbf{fig.}}(\ref{m44}), test particles are a fountain of cold atoms, cooled in an MOT system, launched upward to enter interaction volume {\textbf{fig.}}(\ref{m44}). 
The fountain sets vertically in the gravitational field of earth which we call it gravitational cavity. The missile shape of the setup is designed to function similar to gravitational lenses for matter wave, enhance the deflect angle and therefore the readout resolution. In this setup the turning point act as the cavity mirror. By reading out the displacement on the readout screen, we can learn about the nature of the gravitational potential of the source particle in an inverse scattering manner. 

To optimise the interaction parameters, the mass-ratio between the source and the probe ${M}/{m_{t}}$ is chosen to maximise the readout resolution, minimise the decoherence and more importantly provide the optimal balance between gravitational effect and Casimir-Polder forces. Assuming the objectivity of quantum superposition, as an ansatz, we predict a hypothetical correction in the dynamics of a test particle caused by the spatial superposition of the of a centre of mass of the $M$.

The feasibility study of such an experiment is the subject of an independent research project. Such a project have to estimate the parameter space for a gravitational scattering experiment in nano-scale and micro-distances.
Other relevant parameters, including the interaction time window, decoherence time, external and internal noises, are to be evaluated. For example, {\textbf{fig.}}(\ref{m43}.a)  shows a primary estimation of the gravitational acceleration in the scattering scheme. The main challenge in conducting such a scattering experiment is to fight the competitive effect of Casimir-Polder force at micro-distances. By utilising the proper coating and shielding methods, and also optimising the mass-ratio between source and test particles, we can significantly improve the efficiency of the setup, {\textbf{fig.}}(\ref{m43}.b) shows an example, as a proof of principle, that such an optimisation is possible.

We expect a quantum correction to shows up in the statistical properties of the expected experimental outcomes, caused by the quantum axiom of superposition and uncertainty principle, which govern the dynamics of the centre of mass of $M$. More importantly, the expected outcome of this study verifies the low energy limit of hypothetical full quantum gravity where the gravitating mass is in the quantum regime of motion. Any experimental observation of deviation from linearised gravity would be the first direct quantitative evidence of non-classical features of gravity.

\subsection {Experimental Challenges and Prospective}
\begin{figure}[H]
\begin{center}
a)\begin{subfigure}
       {\includegraphics[width=0.3\textwidth]{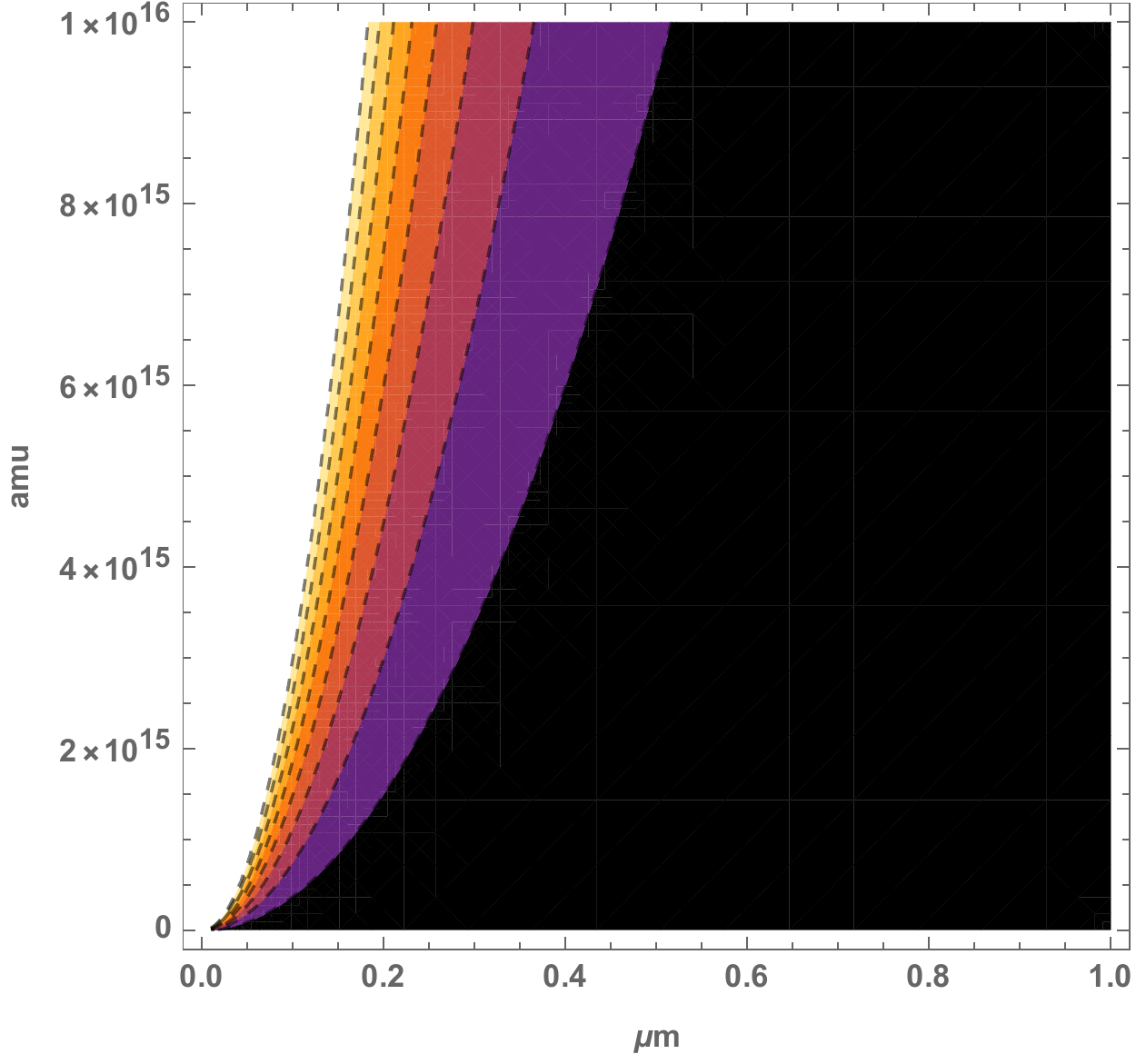}}
       {\includegraphics[width=0.09\textwidth]{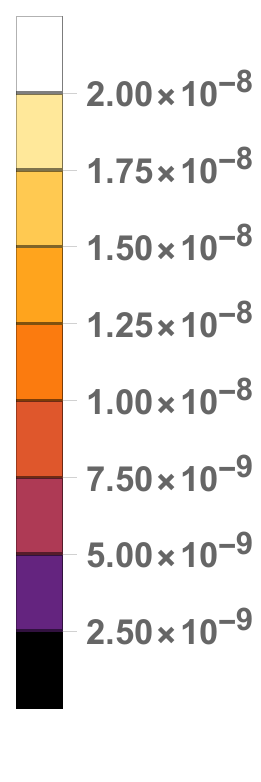}}
\end{subfigure}
b)\begin{subfigure}
     {\includegraphics[width=0.5\textwidth]{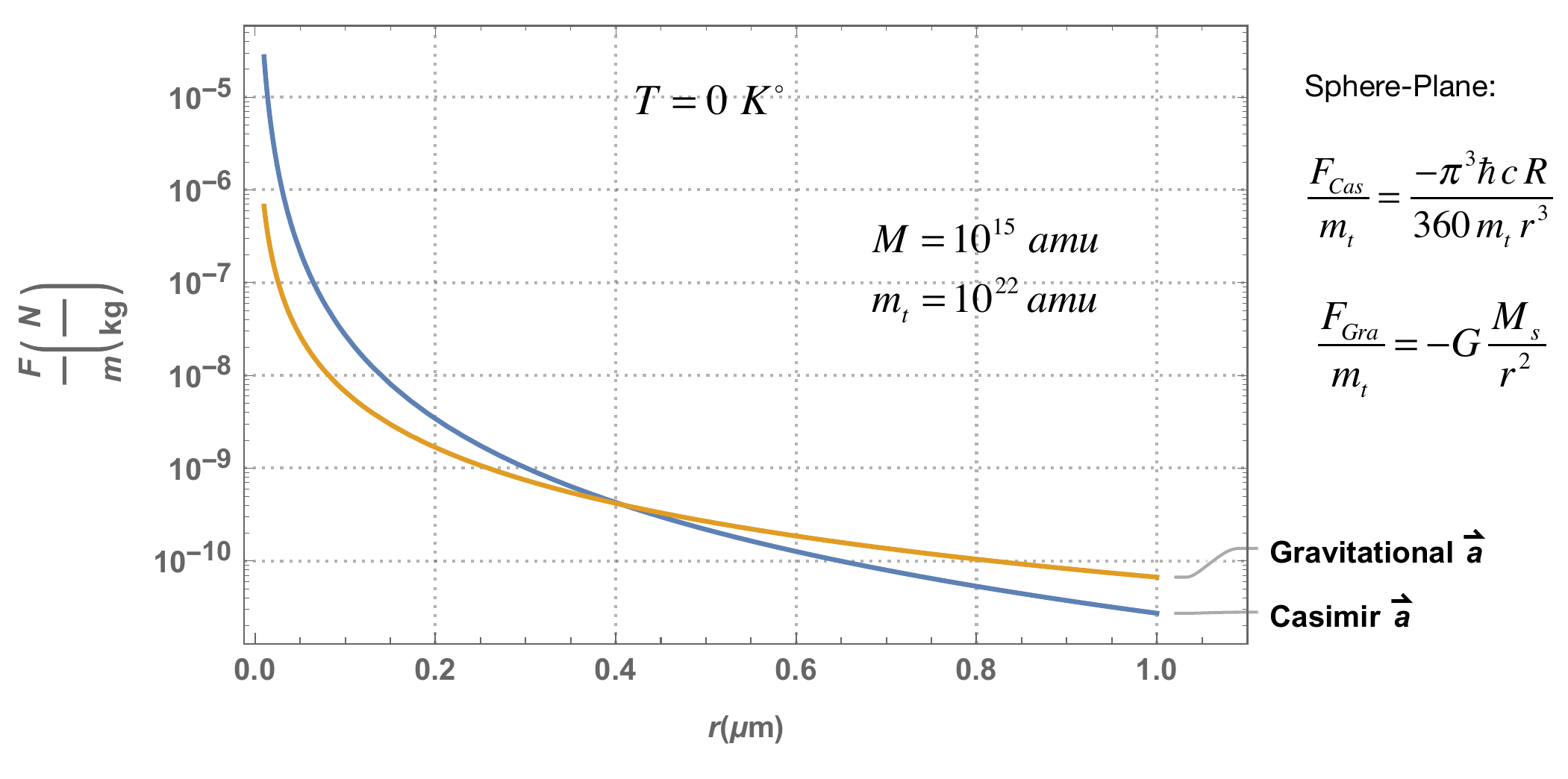}}
\end{subfigure}
\caption{\textbf{a)} The gradual column on the right shows the gravitational acceleration caused by a nanoparticle of different masses (in atomic mass unit) and different impact parameters (distance in $\mu m$) shown on the left graph. \textbf{b)} An example of optimisation: Logarithmic scaling graph is shown to compare the order of magnitude of acceleration from the Casimir-Polder effect to the gravitational field strength (in this example, the source particle $M$ which is going under the spatial superposition is much smaller than the test mass $m_{t}$.) We assumed the maximum value of the Casimir-Polder force (sphere-plane) to be on the safe side. Gravitational acceleration depends only on the mass $M$ of the source particle, while the Casimir-type acceleration depends on the mass of the test particle $m_{t}$. }
\label{m43}
\end{center}
\end{figure}

The most significant challenges for designing a gravitational scattering experiment would be the weakness of gravity and the strength of Casimir-Polder forces. 
An early-stage feasibility study, {\textbf{fig.}}(\ref{m43}.a) and {\textbf{fig.}}(\ref{m43}.b) , shows that with the current technologies conducting such an experiment is not out of access. 
Recent developments in nano-technology and micro-mechanical systems demand a reliable control of Casimir forces; therefore the field has much to offer for filtering or distinguishing Casimir-Polders from the background noise \cite{rodriguez2011casimir}. 
On the other hand, the new generation of force sensing and the latest achievements in precise quantum control and trapping techniques promise the access to a new regime of physics in which the gravitational behaviour of the quantum matter would be experimentally assessable \cite{adelberger2003tests,ulbricht2018testing,geraci2008improved,ranjit2015attonewton}. 
The following is a few suggested initial steps in a feasibility study for testing gravity in microscale:
\begin{itemize}
\item 
Evaluation of external and internal noises, particularly the Casimir-Polder effect. Utilising the methods of coating and shielding, one should consider minimising the Casimir-type effects and find a parameter space in which achieving outcomes would be feasible. 

\item The gravitational scattering off a nano-object is an extremely small effect. It is essential to optimise the design to increase the resolution of the readout. 

\item Statistical analysis of the correlation function at the readout. The outcomes shall be used to solve an inverse scattering problem which describes the nature of the interacting potential. 
\end{itemize}

By analysing the statistical property of the expected experimental outcomes, we can look for an effective correction to the dynamics of the test particle, caused by the nonclassical effect of source mass being in a spatial superposition state. Such an experiment can give us a deeper understanding of how the axiom of quantum superposition and uncertainty principle are interpolated to the construction of space-time in microscale. 

\subsection {A road map of gravity: from SCS to the quantum gravity}

\begin{figure}
\begin{center}
\hspace*{0 cm}{\includegraphics[width=0.8\textwidth]{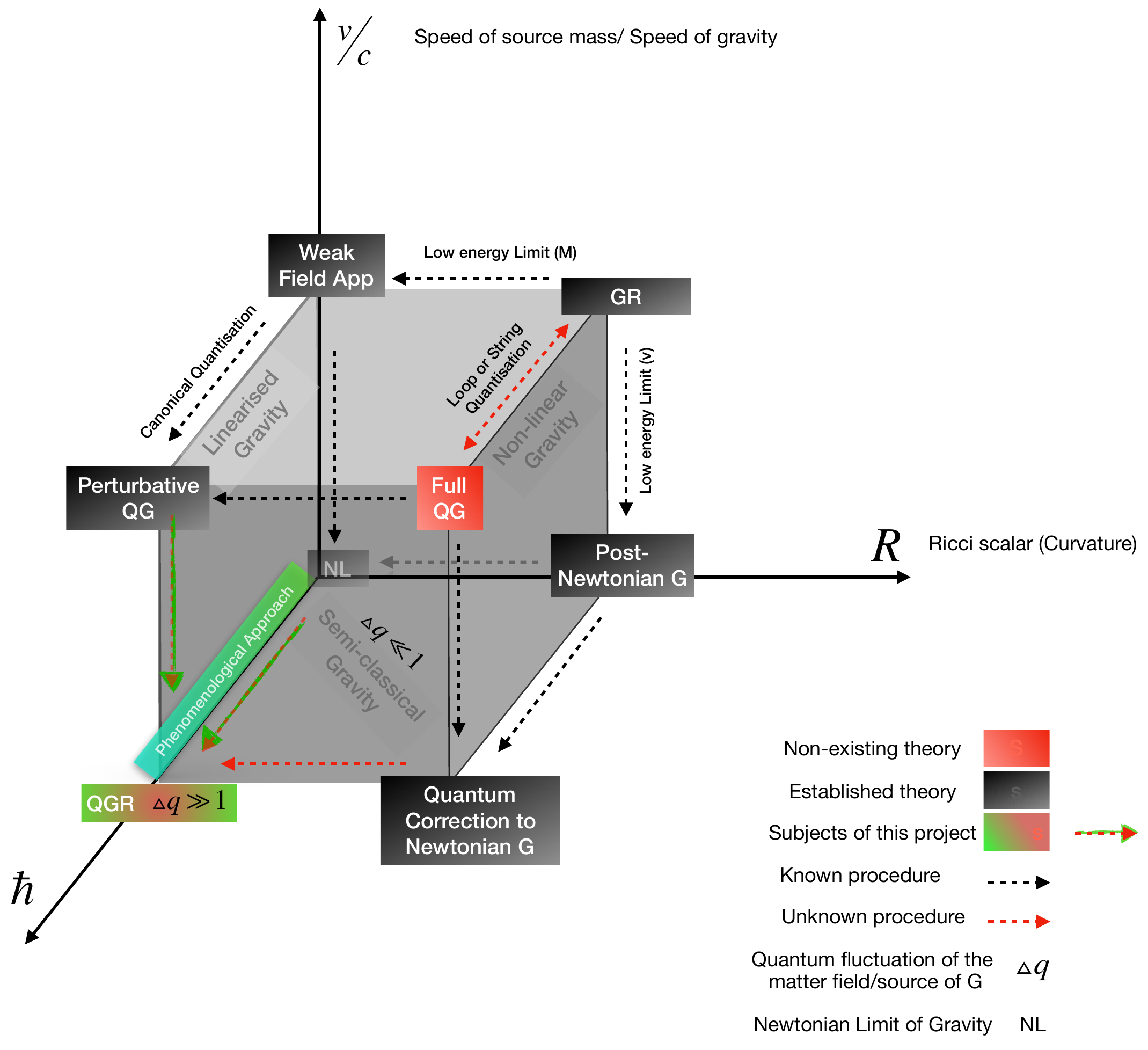}}
\end{center}
\caption{The graph shows how different theories of gravity connected in the limits of various parameters. Three parameters of the gravity are mass (curvature), speed (the ratio between the velocity of source mass and speed of gravity) and Planck constant. When $R << 1$, in the weak field limit, gravity is linear. When $\hbar > 0$, semiclassical gravity is valid except for two particular corners: at the quantum gravity limit where the energy is too high (Planck scale), and at the GQR limit where the fluctuation of the gravitational source is too large. These two corners are not well understood. The latter case (showed in the green shade) is the domain of our study.}
\label{fig:gravity_chart}
\end{figure}

The graph {\textbf{fig.}}(\ref{fig:gravity_chart}) is an illustration to show how the investigation of the GQR can be related or even lead to a better understanding of quantum gravity. In {\textbf{fig.}}(\ref{fig:gravity_chart}), we can see the relation between different theories of gravity connected in the limit of various parameters. Three parameters of the gravity are mass (curvature), speed (the ratio between the velocity of source mass and speed of gravity) and Planck constant. When $R << 1$, in the weak field limit, gravity is linear. When $\hbar > 0$, semiclassical gravity is valid except for two particular corners: at the quantum gravity limit where the energy is too high (Planck scale), and at the GQR limit where the fluctuation of the gravitational source is too large. These two corners are not well understood. The latter case (showed in the green shade) is the domain of our study. The black arrows are showing well-established formalism and methods that fulfil the correspondence principle, where a limiting case of one theory smoothly reaches the other theory. 
As we can see in the graph, red arrows are those areas of ambiguity: going from general theory of gravity to the quantum theory of gravity and vice versa. Surprisingly there is a lack of interest of the corner where all parameters, curvature and velocity, but $\hbar$ are small. This graph better than anything else shows the importance of this research.

\section{Conclusion and Prospective}

We propose an experimental scheme to phenomenologically, explore the feasibility of observing non-classical features of gravity in a linearised, low-energy regime, via a quantum optomechanical experiment. In our experimental scheme, small quantum particles (cold atoms) gravitationally scatter from a massive quantum object (a laser-cooled nano-object), which is prepared in a Spatial Cat State in a double slit setup. By analysing the statistical properties of the expected experimental outcomes, we look for an effective correction to the dynamics of the test particle, caused by the nonclassical effect of a source mass being in a spatial superposition state. 
 Our approach is phenomenological in this sense that we try to adopt the minimal set of assumption on the nature of gravity and mechanism of interaction, to avoid the model-dependence problems.  
We discussed that how deferent reading of quantum mechanics, single versus collective, can play a crucial role in interpreting outcomes of an SCS experiment.
Assuming the validity of Newtonian gravity and the standard non-relativistic quantum mechanics for the table-top range of energy, we have shown that we cannot calculate the gravitational field of an object passing through double-slit. 
Our proposal provides a means to analyse a different aspect of quantum coherence in gravity while avoiding pre-assumption about the nature or form of gravitational interaction. 

{\section*{Acklnowledgment: }} 
The author thanks Ulf Leonhard, Markus Aspelmeyer and Arno Rauschenbeutel who have financed, facilitated and encouraged this research. Exceptional thanks to Dennis R\"atzel, Rainer Kaltenbaek,  Lajos Di{\'o}si, James Millen, Sofia Qvarfort and Itay Griniasty for collaboration, discussion and comments, Karishma Hathlia for her careful and thorough reviews, and Alois Paulin for his support.
\section*{Appendix A.}
\label{appendix:a}

We summarised two different pictures of the concept of the quantum wavefunction which appears in textbooks and literature and both hold with Copenhagen interpretation of quantum mechanics. 
One is closer in semantics to the statistical interpretation of quantum mechanics and implies that wave mechanics is a statistical representation of nature. The second is the single quanta interpretation that believes Schr\"odinger equation should also represent the dynamics of an individual quantum object. The objectivity of the wave function is an unavoidable consequence of the latter picture.

\begin{itemize}
\item {\bf{Wavefunction as the statistical representation of repeated events:}} In this picture, all quantum entities are regarded purely as particles. While any wave-like phenomena are considered to be collective statistical behaviour of many particles. Therefore, while the trajectory of any individual is not predictable, for a large enough sample of particles the distribution over different paths is foreseeable,  due to the rules of quantum mechanics. In this picture, the origin of fringes in a double slit experiment is the uncertainty or intrinsic fluctuation in perpendicular momentum component $p_{y}$ of each particle. Naturally, there is no interference for a single event, but rather the quantum interference is the result of the collective behaviour of a large number of particles or an equally immense number of repetitions \cite{barut1992path}. 

However, this argument contradicts the fact that the Schr\"odinger equation assumes to describe a single, unique particle and {\textbf{eq.}}(\ref{eq:cat_state}) is the answer to that equation. It also fails to explain the mechanism or correlation behind that statistic. Is there any limit on the time gap or spatial interval between sequential particles that breaks the correlation among particles? There is neither a theoretical nor empirical answer to this question. 

Assuming the notion of interference, as a statistical manifestation of a collective correlation, we still have been taken by surprise when the fringes hold even after slowing down the flow of particles to break the instances into single events \cite{juffmann2012real}. {\footnote { By single event here we mean that the spatial interval between particles can reach the order of a centimetre which can be $10^{9}$ more than the diameter of the particle. The time gap between each instance can be arbitrary long.}} 

The best mathematical realisation of such interpretation of the phenomena is the path integral approach \cite{barut1992path}, however applying path integral method to a highly localised object would have a form of underlying semantic of ``ignored'' parameters \cite{barut1993diffraction}. 
Here, we used the term ``ignored'' to distinguish the case from a hidden variable solution like Bohemian mechanics. 
The reading of quantum mechanics can neither quantitively nor qualitatively describe the correlation and the joined probability, between two consecutive particles that arrive at the slits barrier with a substantial spatial and temporal interval.

\item {\bf{Wavefunction as the representation of a single quanta:}} 
Following the de Broglie picture, this reading of quantum mechanics supports a simultaneous coexistence of both wave and particle characteristics for a single quantum object. 
If we suppose that the axioms of quantum mechanics are, or should, be valid for an individual entity, then we may claim that the Spatial Cat State {\textbf{eq.}}(\ref{eq:cat_state}) is the quantum state of each particle. Such an argument would result in the concept of self-interference for an individual particle.  

This interpretation of the SCS {\textbf{eq.}}(\ref{eq:cat_state}) means that the laws of quantum mechanics allow a particle to be in a spatial superposition: a classically forbidden state of being simultaneously in position $x_{1}$ and position $x_{2}$. 
We call such a particle a delocalised quantum object wherein $x_{1} - x_{2}$ is the delocalisation length. Accepting this notion, one can describe the interference pattern of particles observed at the screen.

The mathematical realisation of this interpretation for a double slit is to analytically solve the full Schr\"odinger equation for a single individual particle. To the best knowledge of this author, such a solution has not been published. Beside the path integral approach, the majority of efforts have been given to numerical solutions \cite{turchette1992numerical}. One difficulty arises from the complexity of boundary conditions at the barrier and slits. The other one is to describe a rigid, non-spreading quantum particle by Schr\"odinger equation. 
\end{itemize}

\bibliographystyle{unsrt}
\bibliography{MybibloMM}% Produces the bibliography via BibTeX.
\end{document}